\documentstyle[mprocl,epsf]{article}

\def\be{\begin{equation}}
\def\ee{\end{equation}}
\def\bea{\begin{eqnarray}}
\def\eea{\end{eqnarray}}
\begin{document}

\title{CHARGE-SPIN SEPARATION IN ONE-DIMENSIONAL METALS AND 
INSULATORS~\footnote{To be published in the Proceedings of the Ninth 
International Conference on Recent Progress in Many-Body Theories (Sydney),
ed. by D. Neilson, World Scientific (1998).}}

\author{JOHANNES VOIT}

\address{School of Physics, University of New South Wales, Sydney 2052,
Australia, and Theoretische Physik 1, Universit\"{a}t Bayreuth, 
D-95440 Bayreuth, Germany~\footnote{Present address} \\ 
E-mail: johannes.voit@theo.phy.uni-bayreuth.de}

\maketitle\abstracts{
I discuss origin and possible experimental manifestations of 
charge-spin separation in 1D Luttinger and Luther-Emery liquids,
the latter describing 1D Mott and Peierls insulators and superconductors.
Emphasis is on photoemission where the spectral function generically shows
two dispersing peaks associated with the collective charge and spin 
excitations, and on transport. I analyse the temperature dependences of the
charge and spin conductivities of two organic conductors and conclude that
most likely, charge-spin separation is not realized there and that they
can be described as fluctuating Peierls insulators.}

\section{One-dimensional metals: the Luttinger liquid}
One of the most exciting areas in the condensed-matter many-body problem
is the search for metallic non-Fermi liquid states. This has been sparked
by the unusual normal-state properties of the underdoped high-$T_c$ cuprates
but definite confirmation of a non-Fermi liquid state there is still pending.
This is so mainly because of theoretical difficulties to establish such a 
picture in two dimensions and provide a unifying framework which could tie 
together the multitude of puzzling experimental results. The situation is 
precisely opposite in one dimension (1D): here the Luttinger 
liquid~\cite{Haldane,myrev} has become a paradigm for a non-Fermi liquid metal
but an unambiguous identification of a specific system as a Luttinger 
liquid has not been achieved yet. Unlike 2D, responsible for this is
the wealth of quantitative theoretical information available to date
which puts strong constraints to a Luttinger liquid interpretation of
the otherwise unexplained exotic features of many experiments.

Shortly, a Luttinger liquid can be described as a metal whose elementary 
excitations are not fermionic quasi-particles but collective bosonic charge
and spin fluctuations. This picture emerges 
because the coupling of quasi-particles
to collective modes is extremely strong in 1D~\cite{dr}. 
Quantitatively, one can
pin down two key factors to the breakdown of the Fermi liquid and measure
them by four parameters which completely describe the low-energy physics
in 1D: (i) Singular Peierls-type particle-hole fluctuations which interfere
with superconducting fluctuations, and which lead to non-universal 
power-law scaling of many properties, measured by a coupling constant 
$K_{\rho}$: this parameter parametrizes all scaling laws between the 
various correlation functions. Generically, there is a similar coupling 
constant $K_{\sigma}$ for the spin fluctuations. However, spin-rotation 
invariance requires $K_{\sigma} = 1$ and consequently, this quantity is
often omitted from discussion. (ii) Charge-spin separation, generally 
measured by the 
velocities $v_{\rho,\sigma}$ of the collective modes. 
All gapless 1D quantum systems are 
conjectured to be Luttinger liquids, and no counterexample is known yet
(in theory). 

Candidates for Luttinger liquid correlations are quasi-1D conductors 
auch as (TMTSF)$_2$PF$_6$, i.e. Bechgaard salts~\cite{jer},
semiconductor quantum wires, and 
Quantum Hall edge states~\cite{clark}. Most experiments to date probed
the power-law scaling, i.e. $K_{\rho}$. While generally
consistent with theory, problems concerning the \em values \rm of $K_{\rho}$
remain. An important problem therefore are diagnostic tests of
Luttinger liquid behavior through charge-spin separation. Section 2 addresses 
this issue. Many experiments have been performed
on materials which are 1D but cannot be classified as Luttinger liquids, and 
I will discuss the extent to which charge-spin separation remains visible there
in Section 3. Section 4 discusses charge and spin
transport in 1D organic conductors based on 
fluoranthene and perylene molecules, and to what extent transport
can serve as a diagnostic tool for charge-spin separation.

\section{Charge-spin separation in Luttinger liquids}
To see the origin of charge-spin separation, 
consider a many-body system $H = T + V$ where $T$ is the 
kinetic and $V$ the potential energy. The ground state of $T$ alone is the
filled Fermi sea. If $[T,V] \neq 0$, then the interaction can modify the
ground state by virtual particle-hole excitations (this part of $V$ generates
the nontrivial $K_{\rho} \neq 1$). If the whole $V$ or a part thereof 
commutes with 
$T$, this part cannot change the ground state which remains the Fermi sea,
but it can lift degeneracies of excitations. In 3D, this effect is unimportant
because the spacing of the eigenstates of $T$ scales with the inverse 
linear dimensions of the system while $V$ scales with inverse volume. 
However, both
scales are equal in 1D (A more technical way to express this is
to say that interactions are marginal in 1D.) 
and there is an important reconstruction of the 
spectrum of $H$ in 1D, leading to charge-spin separation. This can be
seen rather easily in a toy problem where an additional particle is put
into the second empty free-particle state above $E_F$ and then $V$ is turned
on: there will be \em two excitation energies \rm associated with the 
additional particle~\cite{myrev}. In general, 
charge-spin separation is an asymptotic property of the low-energy subspace
of $H$.

This toy problem suggests that single-particle properties are important
tests of charge-spin separation. Indeed, the single-electron spectral
function 
\begin{displaymath}
\rho(q,\omega) = (-1/\pi) {\rm Im} G(q+k_F, \omega+\mu)
\end{displaymath}
of a Luttinger
liquid generically, as a hallmark of charge-spin separation, 
has two peaks dispersing with the collective velocities 
$v_{\rho,\sigma}$~\cite{spectral}. The peaks are power-law singularities
governed by a single-particle exponent $\alpha = (1/4) 
\sum_{\nu = \rho, \sigma} (K_{\nu} + K_{\nu}^{-1} -2)$. 
However, for reasons poorly 
understood to date, signatures of these spectral functions have not been found
in photoemission studies of the Bechgaard salts~\cite{fab}. 

\section{1D Peierls and Mott insulators and superconductors}
Photoemission experiments on the quasi-1D conductor
$K_{0.3} Mo O_3$ have shown two dispersing signals~\cite{gweon}, which
disperse with about the Fermi velocity resp.~with $5 v_F$, suggesting
collective character. Such a spectrum is incompatible both with a Fermi
liquid picture and that of a fluctuating Peierls insulator~\cite{lra} 
(a standard model
for the normal state of charge density wave [CDW] systems) 
but consistent with 
charge-spin separation ($v_{\rho} \! > \! v_{\sigma} \! \sim \! v_F$). 
The problem with a Luttinger liquid interpretation 
is that the material undergoes a CDW transition
at 180 K while Luttinger liquids do not have dominant CDW correlations.
As a corrolary, one cannot go directly from a Luttinger liquid to a CDW phase.
Coupling Luttinger liquids to phonons shows that CDW 
correlations can only dominate if a spin gap opens and the system
is in the universality class of the Luther-Emery model with 
$K_{\rho} < 1$~\cite{seoul}. 
Such a 1D spin-gapped Luttinger liquid is the precursor to CDW
formation which would occur at lower temperature as a result of transverse 
coupling. In fact, inverting the Luttinger liquid conjecture, one would
conclude that all 1D systems which are not paramagnetic metals must have
a gap at least in one excitation channel (charge or spin) if not in both.

1D Mott insulators have a charge but no spin gap and also fall into the 
Luther-Emery universality class. Recently, photoemission results have been
published for one such system, $Sr Cu O_2$~\cite{kim}, again with apparently 
two dispersing peaks.
It is not clear, however, what spectral features of a Luttinger liquid
carry over to the Luther-Emery model. Specifically: is charge-spin separation
still visible there?

I have constructed the spectral function of the Luther-Emery
model~\cite{specle}. There are three important differences with respect
to those of a Luttinger liquid: (i) spectral weight is shifted to higher
energies as a result of the opening of a charge or spin gap 
$\Delta_{\rho,\sigma}$; 
(ii) there are strong shadow bands
translating the increased coherence which is at the origin of the gap opening;
(iii) important changes occur in the lineshapes which I now discuss.
In the main band
\begin{figure}[t]
\vspace{0pt}
\epsfxsize=12.5cm
\centerline{\epsfbox{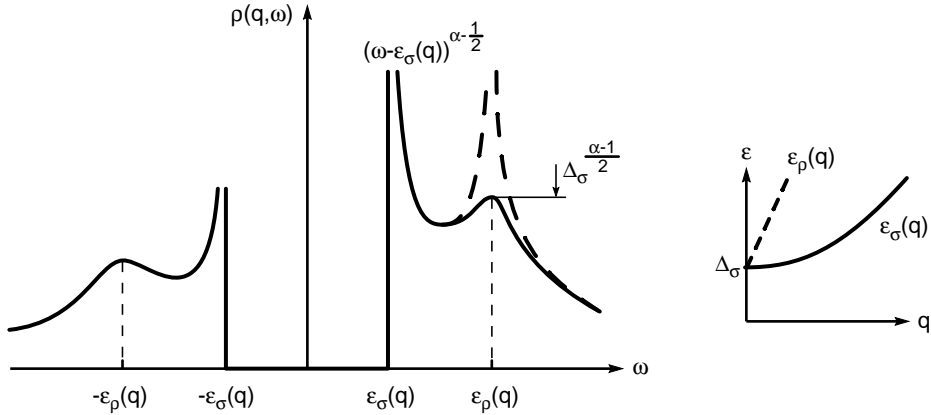}}
%\framebox[55mm]{\rule[-21mm]{0mm}{43mm}}
\caption{Spectral function of the Luther-Emery model, assuming a spin
gap $\Delta_{\sigma}$, for $q>0$ (left).
The main band then is at $\omega > 0$ and the shadow band at $\omega < 0$. 
The right part of the figure shows the
dispersions of both signals in the main band.}
\label{fig:specle}
\end{figure}
[$\omega >(<) 0$ for $q>(<)0$] the number and structure of singularities 
dependes on the relative magnitude of $v_{\rho}$ and $v_{\sigma}$ and
on $K_{\rho}$. For a spin gap and $v_{\rho} > v_{\sigma}$ (as is the case
for repulsive forward scattering and thus dominant CDW correlations), 
I find only \em one \rm true
singularity $[\omega - \{v_{\sigma}^2 q^2 + 
\Delta_{\sigma}^2\}^{1/2}]^{\alpha-1/2}$ 
[$\alpha = (K_{\rho}+K_{\rho}^{-1}-2)/4$ is defined in this way, since
the notion of a $K_{\sigma}$ does not make sense with a spin gap];
the Luttinger liquid's second (charge) singularity is shifted to 
$\Delta_{\sigma}+ v_{\rho}q$ but, importantly, 
\em cut off \rm to a finite maximum
$\sim \! \Delta_{\sigma}^{(\alpha-1)/2}$, cf. Fig.~1. 
If $v_{\sigma} > v_{\rho}$
(generally for attractive forward scattering i.e. a 1D superconductor), 
this becomes a true singularity
$[\omega - \Delta_{\sigma} - v_{\rho} q]^{(\alpha-1)/2}$ while the other
singularity persists. 

In the case of a 1D Mott insulator, 
for the physically relevant case of 
\begin{figure}[t]
\vspace{0pt}
\epsfxsize=12.5cm
\centerline{\epsfbox{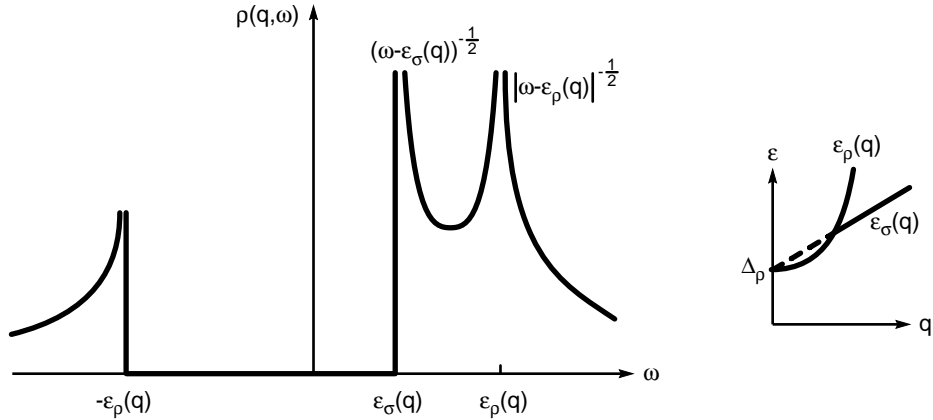}}
%\framebox[55mm]{\rule[-21mm]{0mm}{43mm}}
\caption{Spectral function (left) of a 1D Mott insulator with
$v_{\rho} > v_{\sigma}$ and for wavenumbers beyond the intersection of the
two dispersions. The dispersions of the signals are displayed in the right
part.}
\label{fig:specmott}
\end{figure}
repulsive forward scattering ($v_{\rho} > v_{\sigma}$), one finds
two dispersing singularities in the main band
with universal exponents $-1/2$, Fig.~2. 
However, the two singularities only occur
for wavenumbers larger than a critical one defined by the intersection
of $\varepsilon_{\rho}(q) = (\Delta_{\rho}^2 + v_{\rho}^2 q^2)^{1/2}$ and 
$\varepsilon_{\sigma}(q) = \Delta_{\rho} + v_{\sigma} q$ (this also 
applies to the 1D superconductor: again, simply exchange
all channel labels). A single singularity is observed for $q$ below the
intersection.
Unlike the spin-gap case, the functional form of the
shadow bands here is different: as in the Luttinger liquid~\cite{spectral},
there is no emission at $- \varepsilon_{\sigma}(q)$, and spectral weight
sets in with an inverse-square-root singularity only 
at $-\varepsilon_{\rho}(q)$. In addition, their intensity is suppressed by 
$q$-dependent weight factors.
Numerical calculations are in qualitative agreement~\cite{kim}. For 
$v_{\sigma} > v_{\rho}$ as in the supersymmetric $t-J$ model, there is
only one singularity on the charge dispersion with exponent $-1/2$,
in agreement with a microscopic calculation~\cite{sp}. Except in this latter
case, charge-spin separation therefore is still visible in the spectral
functions of 1D Peierls and Mott insulators. This is consistent with 
photoemission experiments where available~\cite{gweon,kim}.

For $K_{0.3} Mo O_3$ not only photoemission is described correctly by
a Luther-Emery model but also much of the other experimental phenomenlogy
available~\cite{specle,pek}. This statement holds true
\em despite \rm the discussion
by Gweon \em et al.~\cite{gweon}. \rm The Luther-Emery 
model therefore might be a natural starting point for a
description of the low-energy physics of \em some \rm CDW materials such as 
$K_{0.3} Mo O_3$. However, to obtain a good description of the photoemission
properties, one has to assume rather strong electron-electron interactions,
and independent support for such a hypothesis will be required. Another 
problem is that
the spin gap we would obtain from photoemission is much bigger than the one
from susceptibility, and the reasons for this discrepancy are not understood.
The magnitude of this (pseudo)gap, however, 
is also a problem for all competing attempts
to describe these experiments. 

\section{Spin transport and charge spin separation in 1D conductors}
With the ambiguitites and the critical comments by others on the     
evidence in favor of charge-spin separation provided by photoemission,
it is clearly of importance to search for alternative diagnostic tools,
if possible less surface sensitive. I now take up an important suggestion
by Q.~Si~\cite{qimiao} that with
charge-spin separation, and under conditions appropriate for the 2D 
high-$T_c$ superconductors, the charge and spin conductivities of an
itinerant electron system would have different temperature dependences. 
Specifically, I ask if similar correlations can be found in 1D systems. 
We first look at available experiments.

Both conductivities have been measured by others~\cite{fa,pe} 
in the quasi-1D CDW systems (FA)$_2PF_6$ and (PE)$_2PF_6$ (FA = fluoranthene,
PE = perylene). These materials are distinguished by very narrow ESR lines.
Using the Einstein relation $\sigma_{\rm spin}
= \chi D$ with $\chi$ the susceptibility and $D$ the spin diffusion constant,
the spin conductivity can be derived from $D(T)$ measured by ESR. Using the
meanwhile published data of Wokrina \em et al.~\cite{pe}, \rm
the temperature dependences of the charge and spin conductivitites, normalized
to their values at 150 K (I have preferred this temperature for normalization
to, say, room temperature because of the small error bars on the spin 
conductivtiy at 150 K), are obtained as given in Fig.~3. I have reported
a similar analysis for (FA)$_2PF_6$ at a recent conference~\cite{pek}.
\begin{figure}[t]
\vspace{0pt}
\epsfxsize=11.5cm
\centerline{\epsfbox{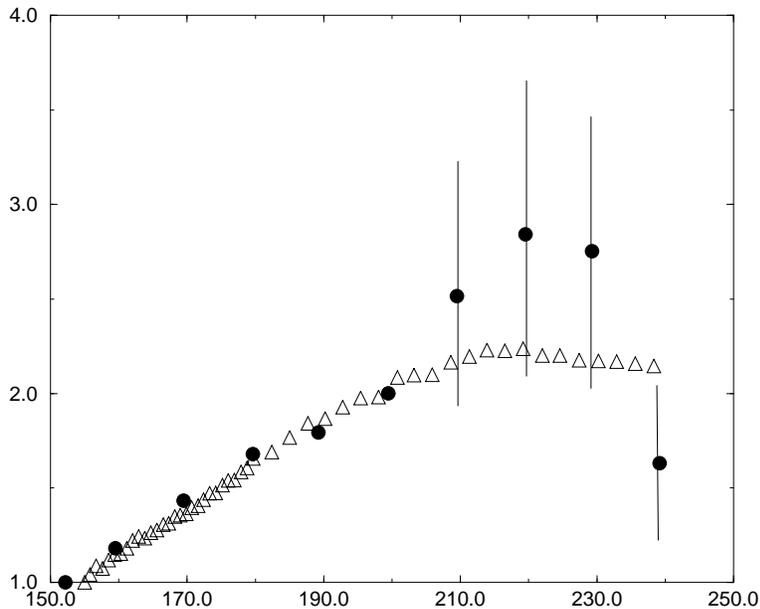}}
%\framebox[55mm]{\rule[-21mm]{0mm}{43mm}}
\caption{Normalized charge (triangles) and spin conductivities (dots) of 
(perylene)$_2PF_6$ vs. temperature, 
from [17].}
\label{fig:conduc}
\end{figure}
Clearly, within the error bars, charge and spin conductivity have the same
temperature dependence.

Is this evidence against 
charge-spin separation? Not necessarily. The idea behind Si's suggestion
is that charge-spin separation, most fundamentally, is a property of
the low-energy sector of the Hilbert space which takes up a product structure.
It is then possible to have scattering processes active
entirely within one of the two factor spaces. In this case, one obtains
the different temperature dependences of charge and spin conductivities.
However, this need not be so: the dissipation process may well couple 
charge and spin degrees of freedom despite charge-spin separation in the
system. This possibility, of course, was known to Si and 
is relevant, e.g., for the process producing the
subdominant temperature dependence in his gauge model 
discussion~\cite{qimiao}. 
To illustrate the point, one can also look at different scattering processes
for the excitations of a Luttinger liquid: (i) electron-electron scattering 
in systems
with even commensurability ($\rho = N_{\rm el}/N_{\rm site} = r/s$ with 
$r,s$ integer, $s$ even) in general produces different temperature 
dependences, as in Si's Luttinger example; 
(ii) with odd commensurability ($s$ odd), similar or dissimilar
temperature dependences for charge and spin conductivities generally depend 
on details of the interactions (through the values of 
$K_{\rho}$); (iii) electron-phonon
and electron-impurity scattering produce identical temperature dependences.
For (FA)$_2PF_6$ and (PE)$_2PF_6$ we thus face the following alternatives:
(1) no charge-spin separation; (2) transport dominated by a scattering 
process involving both charge and spin, independent of charge-spin separation;
(3) both. I now show that most likely, (3) applies. This is different from
Si's situation: in his analysis, the known linear temperature dependence
of the (charge) resistance of the high-$T_c$ copper oxides together with some 
additional information so strongly constrain their description that a 
powerful link between charge-spin separation and the temperature dependences
of the charge and spin conductivities can be made there.

Here, the very similar temperature dependences of charge and 
spin conductivities
strongly suggest similar $T$-dependence of both $D$ and $\chi$ for charge
and spin. Else, differences in the $T$-dependences of charge and spin 
diffusion constants would have to be offset precisely by opposite differences
in the susceptibilities which is unplausible (though impossible to 
exclude strictly). 
The spin diffusion constants provided by ESR generally have weak 
or no temperature 
dependence~\cite{fa,pe}. We then infer that the scattering is 
essentially elastic both for charges and spins, as we expect for
electron-impurity scattering. Organic conductors generally being very pure,
the most natural origin of the ``impurities'' are precursor fluctuations
of the CDW phase formed above the Peierls temperature. Such fluctuations are 
particularly strong in 1D and have been described quantitatively by a
Ginzburg-Landau theory, and their influence on the electronic properties 
is treated as correlated but static disorder~\cite{lra}. One can then
compute the charge and spin conductivities in this model using memory 
functions, and obtains
$\sigma_{\rm charge}(T) \propto \sigma_{\rm spin}(T) \sim \xi(T) \langle
\Psi^2 \rangle_T$ where $\xi$ is the temperature-dependent correlation
length and $\langle \Psi^2 \rangle_T$ are the mean squared fluctuations
of the order parameter $\Psi$. These quantities can be calculated 
exactly for a Ginzburg-Landau model~\cite{ssf}. For $T>T_P$, 
we systematically obtain a concave
temperature dependence quite similar to that observed experimentally, Fig.~3.
Within this model, the strict proportionality of charge and spin 
conductivities is due to its treatment of the CDW precursor fluctuations
as correlated impurities. Moreover, its general neglect of electron-electron
interactions from the outset eliminates any possibility of charge-spin
separation. 

On can also calculate the spectral functions of a fluctuating Peierls 
insulator, Fig.~4. 
\begin{figure}[t]
\vspace{0pt}
\epsfxsize=11.5cm
\centerline{\epsfbox{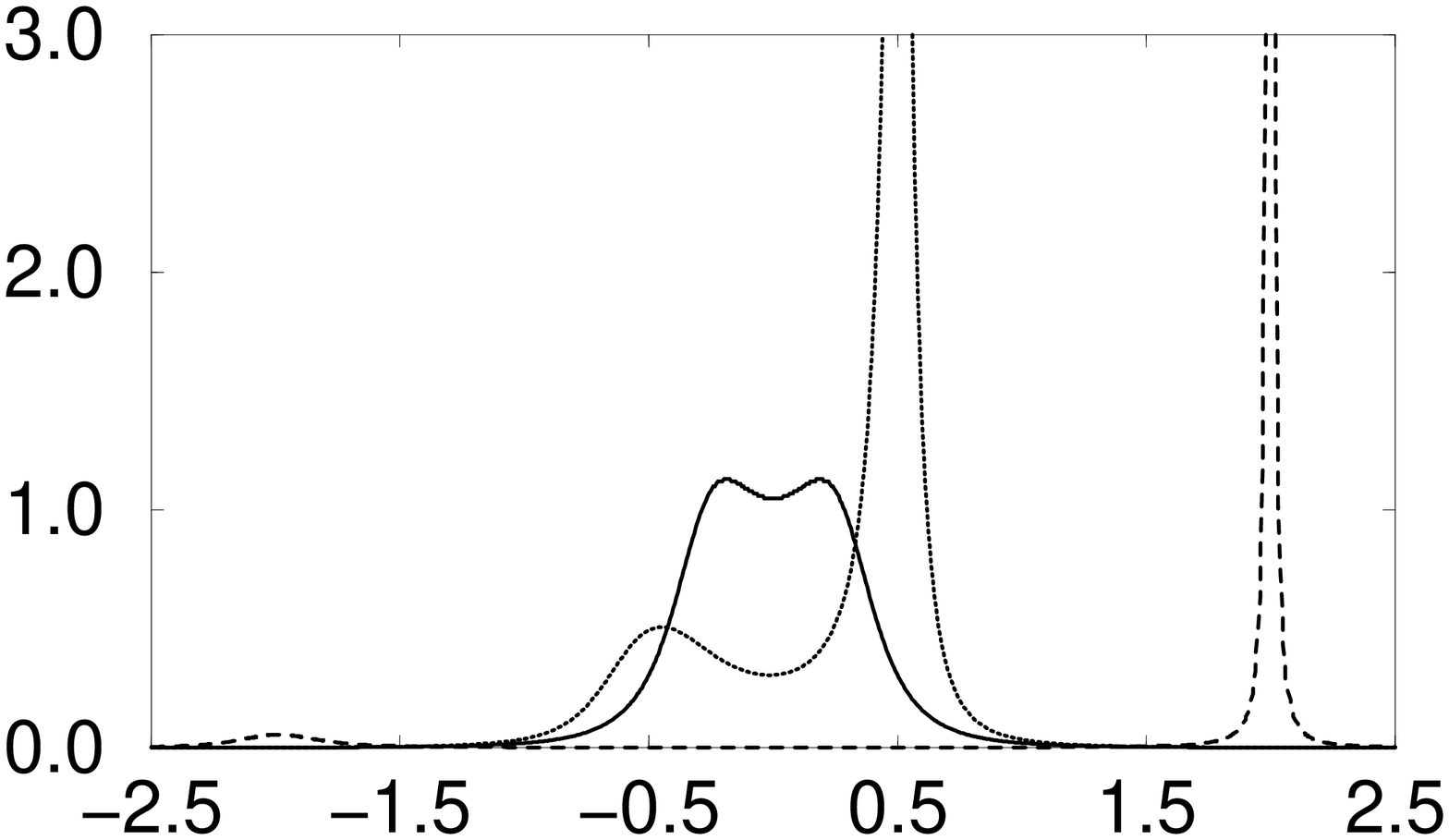}}
%\framebox[55mm]{\rule[-21mm]{0mm}{43mm}}
\caption{Spectral function $\rho(q,\omega)$ vs.~$\omega$
of the Lee-Rice-Anderson model$^{8}$ 
of a fluctuating Peierls insulator: $q=0$ (solid line), $q=0.5$ (dotted line),
$q=2$ (dashed line). }
\label{fig:lrafig}
\end{figure}
While they have only one dispersing signal deep in the range of
(un)occupied states, they are distinctly non-Fermi liquid in shape: they 
broaden as $(q, \omega) \rightarrow 0$ and develop strong shadow bands 
translating the incipient CDW order. 
Both for this model and for the
Luttinger liquid, we are thus in a position to correlate transport and
spectroscopy. Such a correlation in principle also exists in a Fermi 
liquid where transport in described by a Boltzmann equation~\cite{dr} and the
spectral function ideally is dominated by a quasi-particle peak whose
shape, in simple cases, is known.

Finally, it is clear that the model of a fluctuating Peierls insulator 
cannot apply to $K_{0.3}Mo O_3$ (quite independent of what a measurement
of spin conductivity would give): it cannot produce a conductivity 
which could describe the experimental variation $\sigma_{\rm charge}(T)
\sim T^{-1}$ observed for $T > T_P$. 
The difference between $K_{0.3} Mo O_3$ on one side and
(FA)$_2PF_6$ and (PE)$_2PF_6$ on the other is most likely due to the
different electron-phonon coupling which could be comparable to
the electronic interaction in $K_{0.3}MoO_3$ but much stronger in
(FA)$_2(PF)_6$ and (PE)$_2PF_6$.

\vspace*{-2pt}

\section*{Acknowledgments}
I wish to thank Q. Si for many discussions on his work and T. Wokrina and
E. Dormann for sending their data~\cite{pe} prior to publication. 
My work is supported by a Heisenberg fellowship and grant 
SFB 279-B4 of Deutsche Forschungsgemeinschaft, and by the Gordon 
Godfrey Bequest at UNSW.

\eject

\end{document}